# Analysis of the Potential Field and Equilibrium Points of Irregular-shaped Minor Celestial Bodies

Xianyu Wang[1], Yu Jiang[1,2], Shengping Gong[1]

1. School of Aerospace Engineering, Tsinghua University, Beijing 100084, China
2. State Key Laboratory of Astronautic Dynamics, Xi'an Satellite Control Center, Xi'an 710043, China

X. Wang (✉) e-mail: wxy0152@gmail.com (corresponding author)
Y. Jiang (✉) e-mail: jiangyu_xian_china@163.com (corresponding author)

**Abstract.** The equilibrium points of the gravitational potential field of minor celestial bodies, including asteroids, comets, and irregular satellites of planets, are studied. In order to understand better the orbital dynamics of massless particles moving near celestial minor bodies and their internal structure, both internal and external equilibrium points of the potential field of the body are analyzed. In this paper, the location and stability of the equilibrium points of 23 minor celestial bodies are presented. In addition, the contour plots of the gravitational effective potential of these minor bodies are used to point out the differences between them. Furthermore, stability and topological classifications of equilibrium points are discussed, which clearly illustrate the topological structure near the equilibrium points and help to have an insight into the orbital dynamics around the irregular-shaped minor celestial bodies. The results obtained here show that there is at least one equilibrium point in the potential field of a minor celestial body, and the number of equilibrium points could be one, five, seven, and nine, which are all odd integers. It is found that for some irregular-shaped celestial bodies, there are more than four equilibrium points outside the bodies while for some others there are no external equilibrium points. If a celestial body has one equilibrium point inside the body, this one is more likely linearly stable.

**Key words**: Minor bodies; Asteroids; Comets; Equilibrium points; Stability;

## 1. Introduction

During the past few decades, developments in radar shape modeling and direct imaging have made it possible to realize the complexity of the orbital dynamics around irregular-shaped minor celestial bodies; consistently, the topic has become the focus of considerable interest. In this paper we focus on characterizing the dynamical environment of irregular-shaped minor bodies by means of studying the existence, number and stability of equilibrium points in the gravitational potential field of highly irregular-shaped asteroids, comets and satellites of planets. A better understanding of the stability and topological classification of the equilibrium points can help in providing a basis to select the best reconnaissance orbit around asteroids and comets which is essential to the success of space probes sent to study small celestial bodies.

Reliable results have been obtained by previous research devoted to investigate the dynamical environment in the vicinity of asteroids and comets of irregular shape. Elipe and Lara (2003) modeled the asteroid 433 Eros by a finite straight segment and found four equilibrium points around it. Scheeres et al. (2004) used the shape and rotating model from radar imaging data and studied the dynamical environment associated with asteroid 25143 Itokawa; four equilibrium points were found and the coordinates were given in the body-fixed coordinate frame. Blesa (2006) calculated the potential field of two simple-shaped bodies, which included a triangular plate and a square plate. Mondelo et al. (2010) found four equilibrium points around asteroid 4

Vesta and pointed out that there are usually four equilibrium points in the case of an irregular gravitational potential. Liu et al. (2011) calculated the potential field of a rotating cube and found eight equilibrium points. Furthermore, asteroids 216 Kleopatra, 1620 Geographos, 4769 Castalia and 6489 Golevka have been found to have four equilibrium points (Yu and Baoyin 2012; Jiang et al. 2014) while asteroid 1580 Betulia has six equilibrium points and comet 67P/ Churyumov-Gerasimenko has four equilibrium points (Scheeres 2012).

The stability of equilibrium points has a great influence on the dynamical behavior around equilibrium points (Jiang et al. 2014). For the rotating segment, the two collinear equilibrium points are always linearly unstable while the stability of the two isosceles equilibrium points depends on the parameter of the rotating segment (Elipe and Riaguas 2003). For the rotating cube, four equilibrium points are unstable while the other four are linearly stable (Liu et al. 2011). Using precise radar data to model asteroids, it has been established that all of the four equilibrium points outside the asteroids 216 Kleopatra, 1620 Geographos and 4769 Castalia are unstable (Yu and Baoyin 2012; Jiang et al. 2014); while it has been found that two equilibrium points around asteroids 4 Vesta and 6489 Golevka are unstable, the other two being linearly stable (Mondelo et al. 2010; Jiang et al. 2014). Previous research was mainly concerned about the equilibrium points outside of the celestial minor body, neglecting the inner equilibrium points. This is reasonable for planets because there is only one internal equilibrium point and it is located almost in the center of the body. But for the irregular-shaped celestial bodies, the situation can be more complex. The nature of the equilibrium points inside a minor celestial body and their stability may provide information on its internal structure and stresses. This information can be used to study the failure mode of the object. Small asteroids are thought to be gravitational aggregates of loosely consolidated material (e.g. Richardson et al. 2002). The internal dynamical environment may help to investigate the structure of a minor celestial body.

In this paper, we study the number of equilibrium points of the gravitational potential of fifteen asteroids, three comets and five irregular-shaped moonlets of planets. It is found that there exists equilibrium points inside all the celestial minor bodies analyzed here (see Table 1). Besides, for some irregular-shaped celestial bodies, there are more than four equilibrium points outside the celestial body whereas for others there is no outer equilibrium point. For example, there are seven equilibrium points in the potential field of asteroid 216 Kleopatra, four of them are outside the body while the other three are inside it; there is only one equilibrium point in the potential field of asteroid 1998 $KY_{26}$, which is inside the body; there are nine equilibrium points in the potential field of asteroid 101955 Bennu, eight of them are outside the body while only one is inside it. Our results suggest that there is at least one equilibrium point in the gravitational potential of an irregular-shaped minor body; the majority of irregular-shaped minor bodies have five equilibrium points but there are cases that include one, seven or nine.

Stability and topological classifications of equilibrium points are also discussed. It is found that if the minor body has only one inner equilibrium point, such a point is likely to be linearly stable. According to Jiang et al. (2014), the topological structure of the equilibrium point can be classified into five cases. The topological classification of the equilibrium points reveals that all of them belong to one of the Cases 1, 2, or 5 defined in Jiang et al. (2014). The asteroids 4 Vesta, 2867 Steins, 6489 Golevka, 52760, the satellites of planets M1 Phobos, N8 Proteus, S9 Phoebe, and the comets 1P/Halley and 9P/Tempel 1 each have three equilibrium points that belong to Case 1. Moreover, equilibrium points which are outside the irregular-shaped minor body, corresponding to Case 1 and 2, have a staggered distribution. On the other hand, for other minor bodies considered here all the equilibrium points belong either to Case 2 or to Case 5 having those which are outside the minor body a staggered distribution.

## 2. Number and Position of Equilibrium Points

Let us consider the motion of a massless particle around a minor body. The effective potential of the particle can be expressed as (Scheeres et al. 1996; Yu & Baoyin 2012)

$$V(\mathbf{r}) = -\frac{1}{2}(\boldsymbol{\omega} \times \mathbf{r})(\boldsymbol{\omega} \times \mathbf{r}) + U(\mathbf{r}), \tag{1}$$

where $\mathbf{r}$ is the body-fixed vector from the center of mass of the body to the particle, $\boldsymbol{\omega}$ is the rotational angular velocity vector of the body relative to the inertial frame, and $U(\mathbf{r})$ is the gravitational potential. The frame of reference used across this paper is the body-fixed frame. The origin is in the barycenter of the minor body. The *x*, *y*, *z* axes correspond to the principal axes of smallest, intermediate, and largest moment of inertia, respectively.

Using the polyhedron method, the gravitational potential can be written as (Werner

and Scheeres 1996)

$$U = \frac{1}{2}G\sigma \sum_{e \in edges} \mathbf{r}_e \cdot \mathbf{E}_e \cdot \mathbf{r}_e \cdot L_e - \frac{1}{2}G\sigma \sum_{f \in faces} \mathbf{r}_f \cdot \mathbf{F}_f \cdot \mathbf{r}_f \cdot \omega_f,$$

where $G=6.67 \times 10^{-11}$ $m^3kg^{-1}s^{-2}$ represents the gravitational constant, $\sigma$ is the density of the polyhedron, $\mathbf{r}_a$ ($a=e, f$) is a body-fixed vector from the field point to any point on an edge (corresponding to subscript $e$) or face (corresponding to subscript $f$), $L_e$ and $\omega_f$ are factors of integration that operate over the space between the field point and edges or faces, and $\mathbf{E}_e$ and $\mathbf{F}_f$ are dyads representing geometric parameters of edges and faces, which are defined in terms of face- and edge-normal vectors.

The equilibrium points satisfy the following condition (Jiang et al. 2014)

$$\frac{\partial V(x,y,z)}{\partial x} = \frac{\partial V(x,y,z)}{\partial y} = \frac{\partial V(x,y,z)}{\partial z} = 0, \qquad (2)$$

where $(x, y, z)$ are the components of $\mathbf{r}$ in the body-fixed coordinate system. Let $(x_L, y_L, z_L)^T$ denote the coordinates of the critical point; the effective potential $V(x, y, z)$ can be written using a Taylor expansion at the equilibrium point $(x_L, y_L, z_L)^T$. As for the polyhedron models of minor bodies, the potential and gravity can be written as a summation form (Werner and Scheeres 1996). So Eq. (2) can be solved numerically.

If we denote

$$\begin{aligned} \xi &= x - x_L \\ \eta &= y - y_L, \\ \zeta &= z - z_L \end{aligned} \quad \begin{aligned} V_{xx} &= \left(\frac{\partial^2 V}{\partial x^2}\right)_L & V_{xy} &= \left(\frac{\partial^2 V}{\partial x \partial y}\right)_L \\ V_{yy} &= \left(\frac{\partial^2 V}{\partial y^2}\right)_L \quad \text{and} \quad V_{yz} &= \left(\frac{\partial^2 V}{\partial y \partial z}\right)_L, \\ V_{zz} &= \left(\frac{\partial^2 V}{\partial z^2}\right)_L & V_{xz} &= \left(\frac{\partial^2 V}{\partial x \partial z}\right)_L \end{aligned} \qquad (3)$$

the linearized equations of motion relative to the equilibrium point can be expressed as (Jiang et al. 2014)

$$\ddot{\xi} - 2\omega\dot{\eta} + V_{xx}\xi + V_{xy}\eta + V_{xz}\zeta = 0$$
$$\ddot{\eta} + 2\omega\dot{\xi} + V_{xy}\xi + V_{yy}\eta + V_{yz}\zeta = 0 \ . \quad (4)$$
$$\ddot{\zeta} + V_{xz}\xi + V_{yz}\eta + V_{zz}\zeta = 0$$

Let us consider the number of equilibrium points that satisfy the following equation

$$\begin{cases} \dfrac{\partial U(\mathbf{r})}{\partial x} = \omega^2 x \\ \dfrac{\partial U(\mathbf{r})}{\partial y} = \omega^2 y \\ \dfrac{\partial U(\mathbf{r})}{\partial z} = 0 \end{cases} . \quad (5)$$

The asymptotic surface of the effective potential $V = V(\mathbf{r})$ is a circular cylindrical surface when $|\mathbf{r}| \to \infty$. This asymptotic surface is given by the equation $V^* = -\dfrac{\omega^2}{2}(x^2 + y^2)$, so the valued field of the effective potential has no lower bound. On the other hand, the upper bound of the effective potential is an equilibrium point, which suggests that there is at least one equilibrium point in the potential field of a minor body.

If we consider minor celestial bodies which have precise radar observation data (Thomas et al. 1996, 1997; Stooke 1997, 2002; Neese 2004; Nolan 2013), by means of solving Eq. (2), we calculated all the equilibrium points outside and inside minor celestial bodies presented in Table 1, finding out that the number of equilibrium points is not a fixed value; the number depends on the actual shape of the body. Table 1 shows the number of equilibrium points in the potential field of some minor bodies. The first fifteen rows in Table 1 correspond to asteroids, eighteenth to twentieth are satellites of planets, and the last three rows correspond to comets. From Table 1, it can be seen that there is only one equilibrium point in the potential field of asteroid 1998 $KY_{26}$, which is inside the body, whereas asteroid 216 Kleopatra has seven equilibrium points, four of them outside the body and the other three inside, and asteroid 101955 Bennu has nine equilibrium points, eight of them outside the body and the other one inside. It is worth noting that although some minor bodies, such as 2063 Bacchus, 4769 Castalia, 25143 Itokawa, and 433 Eros, have a similar dogbone shape, only 216 Kleopatra has three inner equilibrium points. One possible reason is that only 216 Kleopatra has a really elongated neck between the two lumped masses at the ends. It also indicates that 216 Kleopatra is less stable and it may suffer structural failure in the future (Hirabayashi and Scheeres 2014). A deeper study of the relationship between the inner equilibrium points of a minor body and its structural stability is beyond the scope of the research presented here and it will be attempted in a future paper. Regarding the remaining minor bodies presented in Table 1, each one has five equilibrium points; four of them are outside the body while the other one is inside. Moreover, the results in Table 1 indicate that the number of equilibrium points in the potential field of an irregular shaped body is generally an odd number, such as one, five, seven, or nine. Equilibrium points appear in pairs except the one that is located near the center of the minor body.

Previous work on this subject was usually focused on the equilibrium points that are located outside the minor body and paid no attention to the equilibrium points inside. As for the planets or regular-shaped celestial bodies, it seems that there is only one equilibrium point inside, usually located at the center of the body and therefore of little dynamical relevance. However, in the case of irregular-shaped celestial minor bodies, such as asteroids or comets, the inner equilibrium points may be really meaningful given the fact that most of these minor bodies have not yet been explored and their properties, such as surface composition and density, are based on indirect measurement. The sources of error affecting the computation of the density are coming from the indirect methods used to compute the masses of these minor bodies such as the use of short-term gravitational perturbations derived from the analysis of data on single close encounters between asteroids fitting trajectories computed for a variety of assumed masses of the minor body under consideration to the observed path of the other minor body, the use of spectroscopic analysis and radar albedo, the use of long-term gravitational perturbations in the case of masses derived from periodic variations in the relative positions of moonlets locked in stable orbital resonances, the use of spacecraft tracking data for moonlets, asteroids and comets visited by orbiter and flyby missions, and the use of crude computations of the masses of some comets that have been made by estimating nongravitational forces and comparing them with observed orbital changes. For example, the bulk density of 216 Kleopatra is $3.6\pm0.4$ g cm$^{-3}$ and this value has been obtained from spectroscopic analysis and radar albedo measurements that are used to calculate the density; these computations have their own sources of error, which eventually lead to the error in the bulk density (Descamps et al. 2011). The bulk density error of asteroids is also a result of the estimated errors in the micro-porosity of meteorite or rubble-pile analogues (Marchis et al. 2005).

Moreover, there are several methods (Descamps et al. 2011; Marchis et al. 2005) to determine the surface composition of irregular-shaped minor bodies, including such as the use of spectral reflectance data, the use of thermal infrared spectra and thermal radio data, the use of radar reflectivity (using observations either carried out from Earth or from nearby spacecraft), the use of X-ray and gamma-ray fluorescence (using measurements conducted from an orbiter or flyby spacecraft or made by landing a probe on the body's surface), and the use of chemical analysis of surface samples performed on samples brought to Earth by meteorites or spacecraft, or by in situ analysis using spacecraft. The computation errors of the volume of the minor body are coming from the sources of error associated with the above mentioned methods. Therefore, the error in some of the physical properties of the minor bodies may be very large.

The equilibrium points inside the body can help researchers to study the shape evolution of these irregular-shaped minor bodies. Moreover, it has been suggested that small asteroids are gravitational aggregates of loosely consolidated material (Richardson et al. 2002) meaning that their internal structure is uncertain and their inner stress and internal cohesive forces are unknown. The results obtained in this paper can help to investigate the internal structure of an irregular-shaped minor body by means of the knowledge of the position and stability of the internal equilibrium points as a basis on which further research may lead to know the shape and mass distribution of these minor bodies.

*Table 1. Number of equilibrium points in the potential field of irregular-shaped celestial minor bodies*

| Serial number | Minor bodies | Total number of equilibrium points | Outside | Inside |
|---|---|---|---|---|
| 1 | 4 Vesta | 5 | 4 | 1 |
| 2 | 216 Kleopatra | 7 | 4 | 3 |
| 3 | 243 Ida | 5 | 4 | 1 |
| 4 | 433 Eros | 5 | 4 | 1 |
| 5 | 951 Gaspra | 5 | 4 | 1 |
| 6 | 1620 Geographos | 5 | 4 | 1 |
| 7 | 1996 $HW_1$ | 5 | 4 | 1 |
| 8 | 1998 $KY_{26}$ | 1 | 0 | 1 |
| 9 | 2063 Bacchus | 5 | 4 | 1 |
| 10 | 2867 Steins | 5 | 4 | 1 |
| 11 | 4769 Castalia | 5 | 4 | 1 |
| 12 | 6489 Golevka | 5 | 4 | 1 |
| 13 | 25143 Itokawa | 5 | 4 | 1 |
| 14 | 52760 | 5 | 4 | 1 |
| 15 | 101955 Bennu | 9 | 8 | 1 |
| 16 | J5 Amalthea | 5 | 4 | 1 |
| 17 | M1 Phobos | 5 | 4 | 1 |
| 18 | N8 Proteus | 5 | 4 | 1 |
| 19 | S9 Phoebe | 5 | 4 | 1 |
| 20 | S16 Prometheus | 5 | 4 | 1 |
| 21 | 1P/Halley | 5 | 4 | 1 |
| 22 | 9P/Tempel1 | 5 | 4 | 1 |
| 23 | 103P/Hartley2 | 5 | 4 | 1 |

The positions of equilibrium points for them are presented in Table A1 of Appendix 1 and the physical characteristics of these irregular-shape celestial bodies are presented in Table A2 of Appendix 1. The data in Table A1 are determined by the shape, rotation period and density of minor bodies. Most of the minor bodies have not been visited by spacecraft, so the physical characters are not precise values. As for asteroid 951 Gaspra, 1% of density error and 1% of angular velocity can cost 0.4% and 0.8% of position error of equilibrium points, correspondingly (as show in Figure A1 of Appendix 1). It is worth nothing that asteroid 1998 $KY_{26}$ has only one equilibrium point that is located at the center of the body. This is due to the very short sidereal rotation period of asteroid 1998 $KY_{26}$, which is only 10.704 minutes (Ostro et al. 1999). This makes the centrifugal acceleration around it very large in the body-fixed frame compared with the gravitational acceleration. So there is no equilibrium outside the body. Fig. 1 shows the contour plots of the gravitational effective potential and equilibrium points for minor bodies listed in Table 1. The units of the effective potential per unit mass are m$^2$ s$^{-2}$.

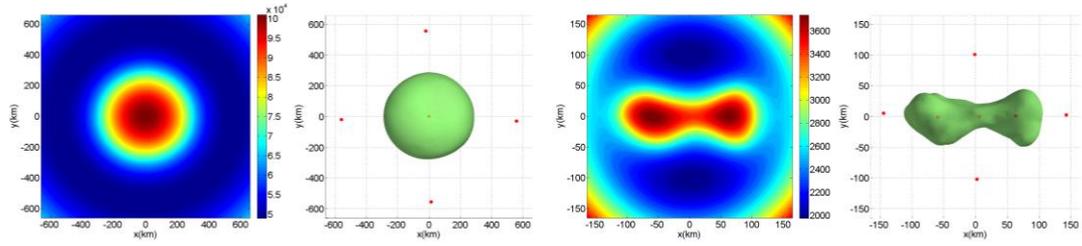
a. 4 Vesta

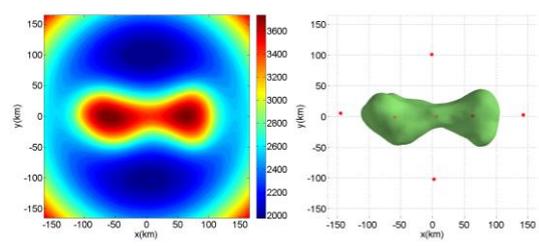
b. 216 Kleopatra

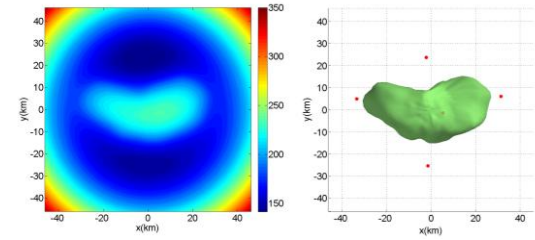
c. 243 Ida

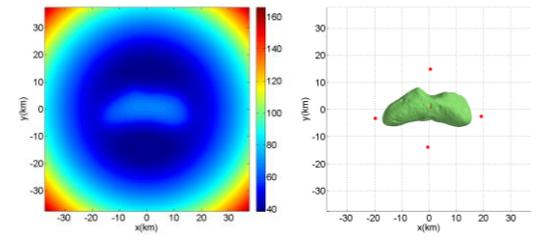
d. 433 Eros

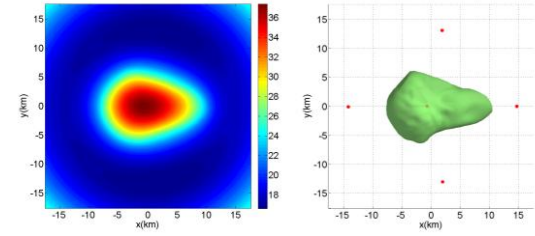
e. 951 Gaspra

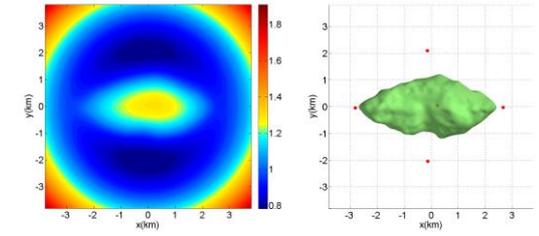
f. 1620 Geographos

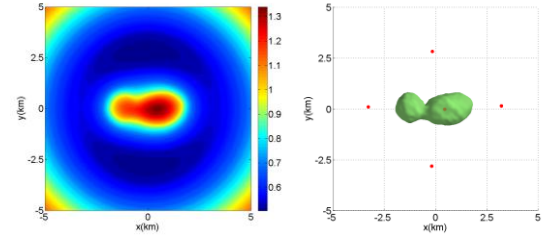
g. 1996 HW$_1$

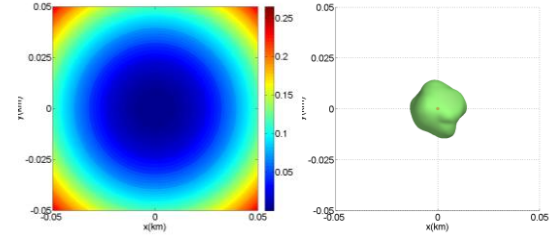
h. 1998 KY$_{26}$

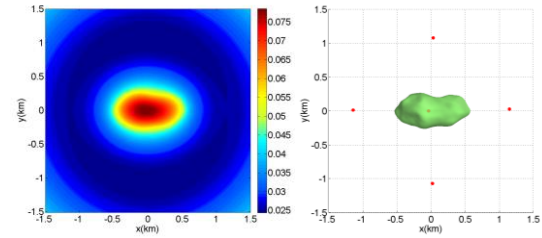
i. 2063 Bacchus

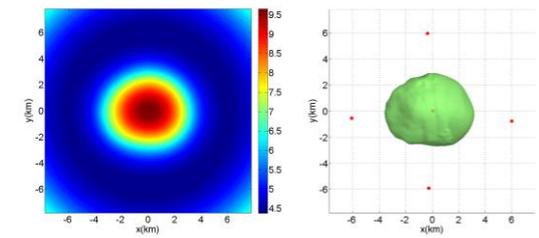
j. 2876 Steins

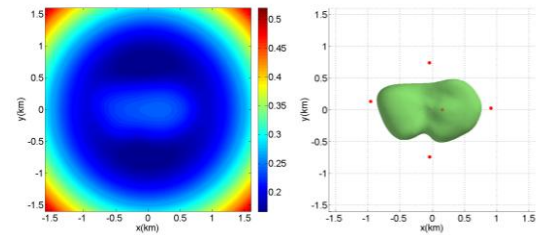
k. 4769 Castalia

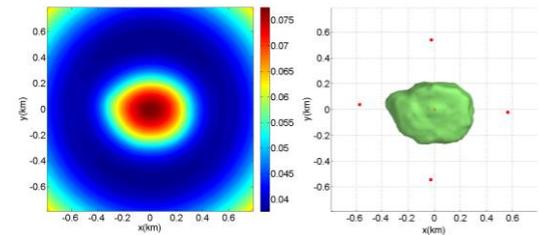
l. 6489 Golevka

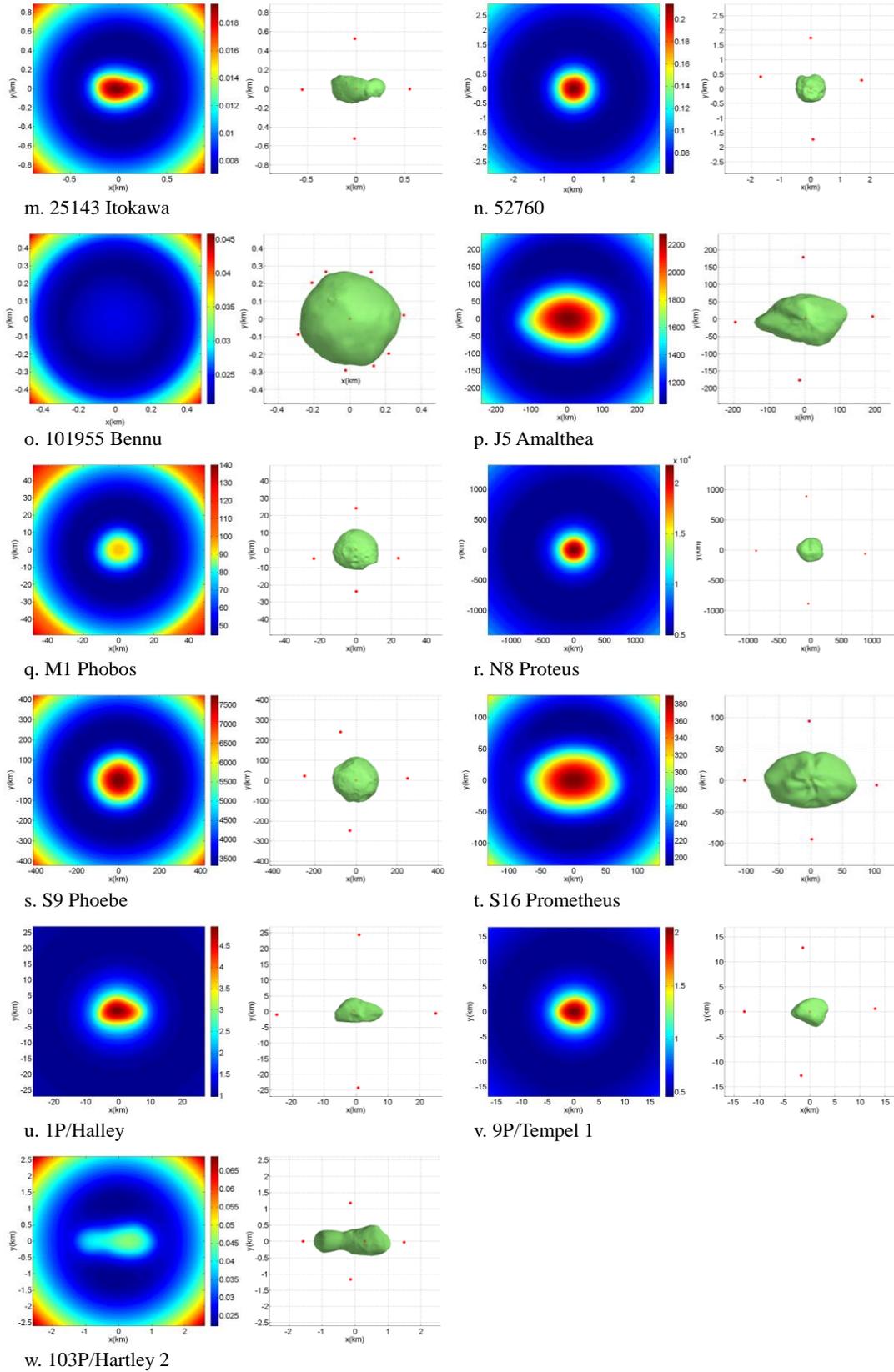

m. 25143 Itokawa  
n. 52760  
o. 101955 Bennu  
p. J5 Amalthea  
q. M1 Phobos  
r. N8 Proteus  
s. S9 Phoebe  
t. S16 Prometheus  
u. 1P/Halley  
v. 9P/Tempel 1  
w. 103P/Hartley 2

Fig. 1. Contour plots of the gravitational effective potential per unit mass and location of equilibrium points for minor bodies. The colour code represents the effective potential per unit mass and the units are $m^2\ s^{-2}$.

## 3. Stability and Topological Classifications of Equilibrium Points

Let **A** be the coefficient matrix of the linearized system,

$$\mathbf{A} = \begin{bmatrix} 0 & 1 \\ -\nabla^2 V & -2\tilde{\boldsymbol{\omega}} \end{bmatrix} \tag{6}$$

where $\tilde{\boldsymbol{\omega}} = \begin{bmatrix} 0 & -\omega & 0 \\ \omega & 0 & 0 \\ 0 & 0 & 0 \end{bmatrix}$, and $\nabla^2 V = \begin{pmatrix} V_{xx} & V_{xy} & V_{xz} \\ V_{xy} & V_{yy} & V_{yz} \\ V_{xz} & V_{yz} & V_{zz} \end{pmatrix}$ is the Hessian matrix of the effective potential. Then Eq. (4) can be written in the following form

$$\dot{\boldsymbol{\varepsilon}} = \mathbf{A}\boldsymbol{\varepsilon} \tag{7}$$

where $\boldsymbol{\varepsilon}$ is the state variable, $\boldsymbol{\varepsilon} = \begin{bmatrix} \xi & \eta & \zeta & \dot{\xi} & \dot{\eta} & \dot{\zeta} \end{bmatrix}^{\mathrm{T}}$.

Let $\mathbf{F}(\mathbf{r}) = \dfrac{\partial V(\mathbf{r})}{\partial \mathbf{r}}$, if the matrix $\nabla^2 V$ is positive definite, the equilibrium point around the celestial body is stable (Jiang et al. 2014).

**Linear classification**: If the Hessian matrix $\nabla^2 V(\boldsymbol{\tau}) = \dfrac{\partial \mathbf{F}(\boldsymbol{\tau})}{\partial \boldsymbol{\tau}}$ has full rank, the equilibrium point is non-degenerate; if the rank of the Hessian matrix $\nabla^2 V(\boldsymbol{\tau})$ is less than 3, the equilibrium point is degenerate.

The Hessian matrix $\nabla^2 V(\boldsymbol{\tau})$ has full rank if and only if $\det(\nabla^2 V(\boldsymbol{\tau})) \neq 0$, in other words, if and only if the Hessian matrix $\nabla^2 V(\boldsymbol{\tau})$ has non-zero eigenvalues. The rank of the Hessian matrix $\nabla^2 V(\boldsymbol{\tau})$ is less than three if and only if $\det(\nabla^2 V(\boldsymbol{\tau})) = 0$, in other words, if and only if the Hessian matrix $\nabla^2 V(\boldsymbol{\tau})$ has zero eigenvalues. The equilibrium points in the potential field of a rotating asteroid can be classified as non-degenerate or degenerate linear.

**Rank classification**: Using the rank of the Hessian matrix $\nabla^2 V(\boldsymbol{\tau})$, the equilibrium points can be classified into $n = 3$ classes, where $n$ is the number of columns of the Hessian matrix. The rank of the Hessian matrix may be 1, 2 or 3, and each value of the rank defines a class of the equilibrium points. Therefore, the equilibrium points in the potential field of a rotating asteroid have three rank classes.

Different classification leads to different laws of motion when we consider the dynamical evolution of a massless particle in the neighborhood of the equilibrium points. Jiang et al. (2014) presented a theorem establishing eight cases of equilibrium points in the potential field of a rotating asteroid that leads to the topological manifold classification of the equilibrium points.

**Topological manifold classification**: Furthermore, for the non-degenerate and non-resonant equilibrium points in the potential field of a celestial body, the topological manifold classification (Jiang et al. 2014) of equilibrium points is presented in Table 2.

Table 2. The topological manifold classification of non-degenerate and non-resonant equilibrium points. C0: Case; C1: Eigenvalues (the imaginary eigenvalues are different); C2: Stability; C3: Number of periodic orbit families around equilibrium points.

| C0 | C1 | C2 | C3 |
|---|---|---|---|
| Case 1 | $\pm i\beta_j \left(\beta_j \in \mathbf{R}^+; j=1,2,3\right)$ | LS | 3 |
| Case 2 | $\pm \alpha_j \left(\alpha_j \in \mathbf{R}^+; j=1\right), \pm i\beta_j \left(\beta_j \in \mathbf{R}^+; j=1,2\right)$ | U | 2 |
| Case 3 | $\pm \alpha_j \left(\alpha_j \in \mathbf{R}^+; j=1,2\right), \pm i\beta_j \left(\beta_j \in \mathbf{R}^+; j=1\right)$ | U | 1 |
| Case 4a | $\pm \alpha_j \left(\alpha_j \in \mathbf{R}^+; j=1\right), \pm \sigma \pm i\tau \left(\sigma, \tau \in \mathbf{R}^+\right)$ | U | 0 |
| Case 4b | $\pm \alpha_j \left(\alpha_j \in \mathbf{R}^+; j=1,2,3\right)$ | U | 0 |
| Case 5 | $\pm \sigma \pm i\tau \left(\sigma, \tau \in \mathbf{R}^+\right), \pm i\beta_j \left(\beta_j \in \mathbf{R}^+; j=1\right)$ | U | 1 |

Each of the irregular-shaped celestial bodies listed in Table 1 is analyzed using Eq. (6) and Eq. (7). All of them are non-degenerate and non-resonant equilibrium points. Therefore, the topological manifold classification is used to analyze the stability of the equilibrium points and the results obtained are presented in Table A3 of Appendix 1. Using the topological manifold classification, we can see from Table A3 that the majority of the equilibrium points located at the center belong to Case 1 except that of the asteroid 216 Kleopatra. This means that the equilibrium point near the center of the body is usually linearly stable, which indicates a stable shape of a celestial body. However, for asteroid 216 Kleopatra, the equilibrium point located at the center of the body is unstable while the other two inner equilibrium points are both linearly stable. This means that its relative tensile strength (RTS) which is close to zero, and typical of weak and porous shattered rubble piles formed by gravitational aggregation, prevents the body from achieving of hydrostatic equilibrium and its dogbone shape suggests a collisional origin where two separate bodies fused together via a gentle collision either from a low-velocity infall of fragments after a disruption event or from tidal decay of a binary system (Magri et al. 2011).

From Tables A1 and A3 in Appendix 1, it can be seen that for each of the minor bodies except the asteroid 216 Kleopatra, the Hessian matrix of the effective potential at the equilibrium point near the barycenter of the body is positive definite. For most of the minor celestial bodies, only the equilibrium point near the barycenter of the body is linearly stable. Nevertheless, for each one of the asteroids 4 Vesta, 2867 Steins, 6489 Golevka, 52760, the satellites of planets M1 Phobos, N8 Proteus, S9

Phoebe, and the comets 1P/Halley and 9P/Tempel 1, there are three linearly stable equilibrium points.

All of the equilibrium points belong to one of the Cases 1, 2, or 5. The equilibrium points in the potential of the asteroids 243 Ida, 433 Eros, 951 Gaspra, 1620 Geographos, 1996 $HW_1$, 2063 Bacchus, 4769 Castalia, 25143 Itokawa, 101955 Bennu, the satellites of planets J5 Amalthea, S16 Prometheus, and the comet 103P/Hartley 2, belong to Case 2 or 5; the number of equilibrium points corresponding to Case 2 is equal to the number of equilibrium points corresponding to Case 5. For these minor bodies, equilibrium points which are outside the body, corresponding to Cases 2 and 5, have a staggered distribution whereas for the asteroids 4 Vesta, 2867 Steins, 6489 Golevka, 52760, the satellites of planets M1 Phobos, N8 Proteus, S9 Phoebe, and the comets 1P/Halley and 9P/Tempel 1, outer equilibrium points that correspond to Cases 1 and 2 also have a staggered distribution.

## 4. Conclusions

In this paper, we have studied the points of equilibrium of the gravitational potential field of irregular-shaped minor celestial bodies such as asteroids, comets, and satellites of planets. The analysis of the results obtained here indicates that there is at least one equilibrium point in the potential field of an irregular-shaped minor body. The number of equilibrium points outside the body is found to be likely either zero, four, or eight but our analysis does not exclude other values. The majority of the irregular-shaped minor bodies studied here have five equilibrium points in their gravitational potential field. Though other values are not excluded, one, seven or nine equilibrium points are also possible. In addition, stability and topological classifications of equilibrium points are analyzed. If the celestial body has only one equilibrium point inside its body, such a point is likely linearly stable. Our analysis shows that, for most of the irregular-shaped bodies, the number of unstable equilibrium points is greater than the number of linearly stable equilibrium points.

Considering the topological classifications of equilibrium points around irregular-shaped minor celestial bodies, all of the equilibrium points belong to one of the Cases 1, 2, or 5. For the asteroids 4 Vesta, 2867 Steins, 6489 Golevka, 52760, the satellites of planets M1 Phobos, N8 Proteus, S9 Phoebe, as well as the comets 1P/Halley and 9P/Tempel 1, the outer equilibrium points, corresponding to Case 1 and 2, have a staggered distribution. For the remaining minor celestial bodies considered here, all the equilibrium points belong to Case 1, 2 or 5, and the outer equilibrium points, corresponding to Cases 2 and 5, have a staggered distribution being the number of equilibrium points that corresponds to Case 2 equals to the one that corresponds to Case 5. The inner equilibrium points are useful for studying the shape evolution and mass distribution of the irregular-shaped minor body whereas the outer equilibrium points can help to understand the orbital dynamics near the celestial minor body.

**Acknowledgements**

We are grateful to the anonymous reviewer for an expeditious review and useful comments. This research was supported by the National Basic Research Program of China (973 Program, 2012CB720000), the State Key Laboratory Foundation of Astronautic Dynamics (No. 2013ADL0202), and the National Natural Science Foundation of China (11372150).

# Appendix 1

Table A1 Positions of the Equilibrium Points for irregular-shaped minor bodies

4 Vesta

| Equilibrium Points | x (km) | y (km) | z (km) |
| --- | --- | --- | --- |
| E1 | 558.306 | -30.0411 | -1.63343 |
| E2 | -20.0372 | 555.912 | -0.639814 |
| E3 | -558.428 | -20.2773 | -0.900566 |
| E4 | 14.0415 | -555.736 | -0.527252 |
| E5 | -0.330992 | -0.047349 | 0.722361 |

216 Kleopatra

| Equilibrium Points | x (km) | y (km) | z (km) |
| --- | --- | --- | --- |
| E1 | 142.852 | 2.44129 | 1.18154 |
| E2 | -1.16383 | 100.740 | -0.545312 |
| E3 | -144.684 | 5.18829 | -0.272463 |
| E4 | 2.22985 | -102.102 | 0.271694 |
| E5 | 63.4440 | 0.827465 | -0.694572 |
| E6 | -59.5425 | -0.969157 | -0.191917 |
| E7 | 6.21924 | -0.198678 | -0.308403 |

243 Ida

| Equilibrium Points | x (km) | y (km) | z (km) |
|---|---|---|---|
| E1 | 31.3969 | 5.96274 | 0.0340299 |
| E2 | -2.16095 | 23.5734 | 0.0975084 |
| E3 | -33.3563 | 4.85067 | -1.08844 |
| E4 | -1.41502 | -25.4128 | -0.378479 |
| E5 | 5.43176 | -1.41369 | -0.144237 |

433 Eros

| Equilibrium Points | x (km) | y (km) | z (km) |
|---|---|---|---|
| E1 | 19.1560 | -2.65188 | 0.142979 |
| E2 | -0.484065 | 14.7247 | -0.0631628 |
| E3 | -19.72858 | -3.38644 | 0.132368 |
| E4 | -0.461655 | -13.9664 | -0.0743819 |
| E5 | 0.549115 | 0.749273 | -0.182043 |

951 Gaspra

| Equilibrium Points | x (km) | y (km) | z (km) |
|---|---|---|---|
| E1 | 14.7323 | -0.0379469 | 0.102217 |
| E2 | 1.90075 | 13.0387 | 0.0112930 |
| E3 | -14.21262 | -0.118726 | 0.0259255 |
| E4 | 1.98791 | -13.0444 | 0.0150812 |
| E5 | -0.692996 | -0.00584606 | -0.0511414 |

1620 Geographos

| Equilibrium Points | x (km) | y (km) | z (km) |
|---|---|---|---|
| E1 | 2.67070 | -0.0398694 | 0.0888751 |
| E2 | -0.142220 | 2.08092 | -0.0220647 |
| E3 | -2.81851 | -0.0557316 | 0.144376 |
| E4 | -0.125676 | -2.04747 | -0.0263415 |
| E5 | 0.228201 | 0.0367998 | -0.0315138 |

1996 $HW_1$

| Equilibrium Points | x (km) | y (km) | z (km) |
|---|---|---|---|
| E1 | 3.21197 | 0.133831 | -0.00232722 |
| E2 | -0.150078 | 2.80789 | 0.000515378 |
| E3 | -3.26866 | 0.0841431 | -0.00103271 |
| E4 | -0.181051 | -2.82605 | 0.000146216 |
| E5 | 0.452595 | -0.0291869 | 0.00302067 |

1998 KY$_{26}$

| Equilibrium Points | x (km) | y (km) | z (km) |
|---|---|---|---|
| E1 | 0.00000007094 | -0.00000032002 | -0.0000109550 |

2063 Bacchus

| Equilibrium Points | x (km) | y (km) | z (km) |
|---|---|---|---|
| E1 | 1.14738 | 0.0227972 | -0.000861348 |
| E2 | 0.0314276 | 1.07239 | 0.000711379 |
| E3 | -1.14129 | 0.00806235 | -0.00141486 |
| E4 | 0.0203102 | -1.07409 | 0.000849894 |
| E5 | -0.0362491 | -0.00393237 | 0.00222295 |

2867 Steins

| Equilibrium Points | x (km) | y (km) | z (km) |
|---|---|---|---|
| E1 | 6.02496 | -0.778495 | 0.0194957 |
| E2 | -0.350453 | 5.93496 | -0.0389495 |
| E3 | -6.09443 | -0.556101 | 0.0108389 |
| E4 | -0.258662 | -5.91204 | -0.0328477 |
| E5 | 0.0465217 | 0.0118791 | 0.00862801 |

4769 Castalia

| Equilibrium Points | x (km) | y (km) | z (km) |
|---|---|---|---|
| E1 | 0.910109 | 0.0228648 | 0.0345927 |
| E2 | -0.0427816 | 0.736033 | 0.00312877 |
| E3 | -0.953021 | 0.128707 | 0.0300658 |
| E4 | -0.0399531 | -0.744131 | 0.00876237 |
| E5 | 0.157955 | -0.00144811 | -0.0129416 |

6489 Golevka

| Equilibrium Points | x (km) | y (km) | z (km) |
|---|---|---|---|
| E1 | 0.564128 | -0.023416 | -0.002882 |
| E2 | -0.571527 | 0.035808 | -0.006081 |
| E3 | -0.021647 | 0.537470 | -0.001060 |
| E4 | -0.026365 | -0.546646 | -0.000182 |
| E5 | 0.002330 | -0.003329 | 0.002198 |

25143 Itokawa

| Equilibrium Points | x (km) | y (km) | z (km) |
|---|---|---|---|
| E1 | 0.554478 | -0.00433107 | -0.000061 |
| E2 | -0.0120059 | 0.523829 | -0.000201 |
| E3 | -0.555624 | -0.0103141 | -0.000274 |
| E4 | -0.0158721 | -0.523204 | 0.000246 |
| E5 | 0.00346405 | 0.00106939 | 0.000105 |

52760

| Equilibrium Points | x (km) | y (km) | z (km) |
|---|---|---|---|
| E1 | 1.71569 | 0.286648 | -0.00123816 |
| E2 | -0.000226269 | 1.73333 | -0.000340622 |
| E3 | -1.69114 | 0.407657 | -0.00137500 |
| E4 | 0.0777576 | -1.73261 | 0.00003270 |
| E5 | 0.000487075 | -0.00200895 | -0.000137221 |

101955 Bennu

| Equilibrium Points | x (km) | y (km) | z (km) |
|---|---|---|---|
| E1 | 0.302254 | 0.0207971 | -0.00325871 |
| E2 | 0.119921 | 0.263907 | -0.00250975 |
| E3 | -0.133380 | 0.265869 | -0.0108739 |
| E4 | -0.211138 | 0.204168 | -0.00979061 |
| E5 | -0.288326 | -0.884228 | -0.00262675 |
| E6 | -0.00230678 | -0.290888 | 0.00159251 |
| E7 | 0.133622 | -0.265726 | -0.00161864 |
| E8 | 0.217825 | -0.196538 | -0.00365738 |
| E9 | 0.000149629 | 0.00020317 | 0.00004969 |

J5 Amalthea

| Equilibrium Points | x (km) | y (km) | z (km) |
|---|---|---|---|
| E1 | 193.626 | 6.99477 | -0.526508 |
| E2 | -3.70510 | 177.930 | 0.503297 |
| E3 | -196.393 | -9.22749 | -1.68477 |
| E4 | -13.9926 | -177.033 | 0.827095 |
| E5 | 2.67129 | 0.497929 | 1.20783 |

M1 Phobos

| Equilibrium Points | x (km) | y (km) | z (km) |
|---|---|---|---|
| E1 | 24.0314 | -4.77124 | -0.163089 |
| E2 | 0.0317206 | 240.5832 | 0.703125 |
| E3 | -23.9533 | -4.99313 | -0.0684951 |
| E4 | 0.103851 | -23.9175 | 0.0955774 |
| E5 | -0.0767803 | 0.0677243 | 0.0811266 |

N8 Proteus

| Equilibrium Points | x (km) | y (km) | z (km) |
|---|---|---|---|
| E1 | 887.221 | -66.3922 | 0.309053 |
| E2 | -71.745.2 | 884.930 | -0.00239582 |
| E3 | -890.219 | -12.8162 | 0.402503 |
| E4 | -40.8827 | -886648 | -0.0538940 |
| E5 | 0.400639 | -0.129694 | -0.522757 |

S9 Phoebe

| Equilibrium Points | x (km) | y (km) | z (km) |
|---|---|---|---|
| E1 | 251.227 | 9.72748 | 0.193940 |
| E2 | -75.5341 | 239.050 | -0.492151 |
| E3 | -250.944 | 21.3226 | 0.470713 |
| E4 | -30.0715 | -248.639 | -0.361233 |
| E5 | 0.148101 | -0.191564 | -0.0134649 |

S16 Prometheus

| Equilibrium Points | x (km) | y (km) | z (km) |
|---|---|---|---|
| E1 | 103.638 | -7.63664 | -0.206733 |
| E2 | -2.44497 | 94.0209 | 0.194146 |
| E3 | -103.999 | 0.106813 | -1.04257 |
| E4 | 1.53394 | -93.7212 | -0.120004 |
| E5 | -0.238211 | 0.351697 | 0.337193 |

1P/Halley

| Equilibrium Points | x (km) | y (km) | z (km) |
|---|---|---|---|
| E1 | 24.9463 | -0.662821 | 0.00372359 |
| E2 | 0.944749 | 24.3406 | -0.00137676 |
| E3 | -24.8561 | -1.05235 | 0.00508783 |
| E4 | 0.674504 | -24.3224 | -0.00007979 |
| E5 | -0.576790 | 0.142134 | 0.00830363 |

9P/Tempel 1

| Equilibrium Points | x (km) | y (km) | z (km) |
|---|---|---|---|
| E1 | 12.9584 | 0.609590 | -0.00574687 |
| E2 | -1.37819 | 12.7883 | -0.00398703 |
| E3 | -12.9853 | 0.0332369 | -0.00416586 |
| E4 | -1.71181 | -12.7459 | -0.00774402 |
| E5 | 0.0080638 | 0.0054987 | 0.0216501 |

103P/Hartley 2

| Equilibrium Points | x (km) | y (km) | z (km) |
|---|---|---|---|
| E1 | 1.48975 | -0.0343398 | 0.00953198 |
| E2 | -0.142516 | 1.17411 | -0.00293228 |
| E3 | -1.58280 | -0.00593462 | -0.00326151 |
| E4 | -0.137522 | -1.17362 | -0.00278246 |
| E5 | 0.297320 | 0.00107410 | -0.00593623 |

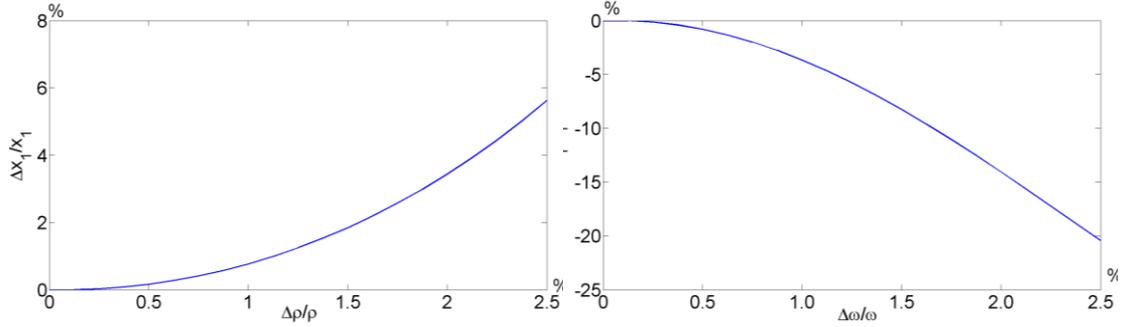

Figure A1 The x component of E1 of 951 Gaspra variation with density and angular velocity

Table A2 Physical properties of irregular-shaped celestial minor bodies

| Serial number | Minor bodies | Bulk density(g/cm$^3$) | Rotation period(h) |
| --- | --- | --- | --- |
| 1 | 4 Vesta[a] | 3.456 | 5.342 |
| 2 | 216 Kleopatra[b1,b2] | 4.27 | 5.385 |
| 3 | 243 Ida[c1,c2] | 2.6 | 4.63 |
| 4 | 433 Eros[d] | 2.67 | 5.27 |
| 5 | 951 Gaspra[e1,e2] | 2.71 | 7.042 |
| 6 | 1620 Geographos[f] | 2.0 | 5.223 |
| 7 | 1996 HW$_1$[g1,g2] | 3.56 | 8.757 |
| 8 | 1998 KY$_{26}$[h] | 2.8 | 0.1784 |
| 9 | 2063 Bacchus[i] | 2.0 | 14.9 |
| 10 | 2867 Steins[j1,j2] | 1.8 | 6.04679 |
| 11 | 4769 Castalia[k1,k2] | 2.1 | 4.094 |
| 12 | 6489 Golevka[l1,l2] | 2.7 | 6.026 |
| 13 | 25143 Itokawa[m1,m2] | 1.95 | 12.132 |
| 14 | 52760[n] | 2.5 | 14.98 |
| 15 | 101955 Bennu[o1,o2] | 0.97 | 4.288 |
| 16 | J5 Amalthea[p1,p2] | 0.857 | 11.9564 |
| 17 | M1 Phobos[q1,q2] | 1.876 | 7.65 |
| 18 | N8 Proteus[r1,r2] | 1.3 | 26.9 |
| 19 | S9 Phoebe[s] | 1.63 | 9.27 |
| 20 | S16 Prometheus[t1,t2] | 0.48 | 14.71 |
| 21 | 1P/Halley[u1,u2] | 0.6 | 52.8 |
| 22 | 9P/Tempel1[v1,v2] | 0.62 | 40.7 |
| 23 | 103P/Hartley2[w] | 0.34 | 18.0 |

Table A3 Topological manifold classifications of the Equilibrium Points around minor bodies. LS: linearly stable; U: unstable;

P: positive definite; N: non-positive definite

4 Vesta

| Equilibrium Points | Case | Stability | $\nabla^2 V$ |
|---|---|---|---|
| E1 | 2 | U | N |
| E2 | 1 | LS | P |
| E3 | 2 | U | N |
| E4 | 1 | LS | P |
| E5 | 1 | LS | P |

216 Kleopatra

| Equilibrium Points | Case | Stability | $\nabla^2 V$ |
|---|---|---|---|
| E1 | 2 | U | N |
| E2 | 5 | U | N |
| E3 | 2 | U | N |
| E4 | 5 | U | N |
| E5 | 1 | LS | P |
| E6 | 1 | LS | P |
| E7 | 2 | U | N |

243 Ida

| Equilibrium Points | Case | Stability | $\nabla^2 V$ |
|---|---|---|---|
| E1 | 2 | U | N |
| E2 | 5 | U | N |
| E3 | 2 | U | N |
| E4 | 5 | U | N |
| E5 | 1 | LS | P |

433 Eros

| Equilibrium Points | Case | Stability | $\nabla^2 V$ |
|---|---|---|---|
| E1 | 2 | U | N |
| E2 | 5 | U | N |
| E3 | 2 | U | N |
| E4 | 5 | U | N |
| E5 | 1 | LS | P |

## 951 Gaspra

| Equilibrium Points | Case | Stability | $\nabla^2 V$ |
|---|---|---|---|
| E1 | 2 | U | N |
| E2 | 5 | U | N |
| E3 | 2 | U | N |
| E4 | 5 | U | N |
| E5 | 1 | LS | P |

## 1620 Geographos

| Equilibrium Points | Case | Stability | $\nabla^2 V$ |
|---|---|---|---|
| E1 | 2 | U | N |
| E2 | 5 | U | N |
| E3 | 2 | U | N |
| E4 | 5 | U | N |
| E5 | 1 | LS | P |

## 1996 HW$_1$

| Equilibrium Points | Case | Stability | $\nabla^2 V$ |
|---|---|---|---|
| E1 | 2 | U | N |
| E2 | 5 | U | N |
| E3 | 2 | U | N |
| E4 | 5 | U | N |
| E5 | 1 | LS | P |

## 1998 KY$_{26}$

| Equilibrium Points | Case | Stability | $\nabla^2 V$ |
|---|---|---|---|
| E1 | 1 | LS | P |

## 2063 Bacchus

| Equilibrium Points | Case | Stability | $\nabla^2 V$ |
|---|---|---|---|
| E1 | 2 | U | N |
| E2 | 5 | U | N |
| E3 | 2 | U | N |
| E4 | 5 | U | N |
| E5 | 1 | LS | P |

## 2867 Steins

| Equilibrium Points | Case | Stability | $\nabla^2 V$ |
|---|---|---|---|
| E1 | 2 | U | N |
| E2 | 1 | LS | P |
| E3 | 2 | U | N |
| E4 | 1 | LS | P |
| E5 | 1 | LS | P |

## 4769 Castalia

| Equilibrium Points | Case | Stability | $\nabla^2 V$ |
|---|---|---|---|
| E1 | 2 | U | N |
| E2 | 5 | U | N |

| Equilibrium Points | Case | Stability | $\nabla^2 V$ |
|---|---|---|---|
| E3 | 2 | U | N |
| E4 | 5 | U | N |
| E5 | 1 | LS | P |

6489 Golevka

| Equilibrium Points | Case | Stability | $\nabla^2 V$ |
|---|---|---|---|
| E1 | 2 | U | N |
| E2 | 1 | LS | N |
| E3 | 2 | U | N |
| E4 | 1 | LS | N |
| E5 | 1 | LS | P |

25143 Itokawa

| Equilibrium Points | Case | Stability | $\nabla^2 V$ |
|---|---|---|---|
| E1 | 2 | U | N |
| E2 | 5 | U | N |
| E3 | 2 | U | N |
| E4 | 5 | U | N |
| E5 | 1 | LS | P |

52760

| Equilibrium Points | Case | Stability | $\nabla^2 V$ |
|---|---|---|---|
| E1 | 2 | U | N |
| E2 | 1 | LS | P |
| E3 | 2 | U | N |
| E4 | 1 | LS | P |
| E5 | 1 | LS | P |

101955 Bennu

| Equilibrium Points | Case | Stability | $\nabla^2 V$ |
|---|---|---|---|
| E1 | 2 | U | N |
| E2 | 5 | U | N |
| E3 | 2 | U | N |
| E4 | 5 | U | N |
| E5 | 2 | U | N |
| E6 | 5 | U | N |
| E7 | 2 | U | N |
| E8 | 5 | U | N |
| E9 | 1 | LS | P |

J5 Amalthea

| Equilibrium Points | Case | Stability | $\nabla^2 V$ |
|---|---|---|---|
| E1 | 2 | U | N |
| E2 | 5 | U | N |
| E3 | 2 | U | N |
| E4 | 5 | U | N |
| E5 | 1 | LS | P |

## M1 Phobos

| Equilibrium Points | Case | Stability | $\nabla^2 V$ |
|:---:|:---:|:---:|:---:|
| E1 | 2 | U | N |
| E2 | 1 | LS | P |
| E3 | 2 | U | N |
| E4 | 1 | LS | P |
| E5 | 1 | LS | P |

## N8 Proteus

| Equilibrium Points | Case | Stability | $\nabla^2 V$ |
|:---:|:---:|:---:|:---:|
| E1 | 2 | U | N |
| E2 | 1 | LS | P |
| E3 | 2 | U | N |
| E4 | 1 | LS | P |
| E5 | 1 | LS | P |

## S9 Phoebe

| Equilibrium Points | Case | Stability | $\nabla^2 V$ |
|:---:|:---:|:---:|:---:|
| E1 | 2 | U | N |
| E2 | 1 | LS | P |
| E3 | 2 | U | N |
| E4 | 1 | LS | P |
| E5 | 1 | LS | P |

## S16 Prometheus

| Equilibrium Points | Case | Stability | $\nabla^2 V$ |
|:---:|:---:|:---:|:---:|
| E1 | 2 | U | N |
| E2 | 5 | U | N |
| E3 | 2 | U | N |
| E4 | 5 | U | N |
| E5 | 1 | LS | P |

## 1P/Halley

| Equilibrium Points | Case | Stability | $\nabla^2 V$ |
|:---:|:---:|:---:|:---:|
| E1 | 2 | U | N |
| E2 | 1 | LS | N |
| E3 | 2 | U | N |
| E4 | 1 | LS | N |
| E5 | 1 | LS | P |

## 9P/Tempel1

| Equilibrium Points | Case | Stability | $\nabla^2 V$ |
|:---:|:---:|:---:|:---:|
| E1 | 2 | U | N |
| E2 | 1 | LS | N |
| E3 | 2 | U | N |
| E4 | 1 | LS | N |
| E5 | 1 | LS | P |

| 103P/Hartley2 | | | |
|---|---|---|---|
| Equilibrium Points | Case | Stability | $\nabla^2 V$ |
| E1 | 2 | U | N |
| E2 | 5 | U | N |
| E3 | 2 | U | N |
| E4 | 5 | U | N |
| E5 | 1 | LS | P |